\documentclass[preprint,aps,prb,showpacs,amsmath,amssymb]{revtex4}
\usepackage{graphics}
\usepackage{epsf,graphicx,pstcol,epsfig,psfrag}
\begin{document}

\title{Measure and collapse of participatory democracy in a two party system.}
\author{Jozef Sznajd}
\affiliation{Institute for Low Temperature and Structure Research, Polish Academy of Sciences, Wroclaw} 
\date{\today}

\begin{abstract}
\emph{Measure what is measurable, and make measurable what is not so
(Galileo Galilei)}. According to the above sentence we do not ask why we need to measure democracy but if it is possible to measure something which is not unequivocally defined. Although, it is unlikely a final agreement on the definition of democracy, the idea that it is a form of governance based on collective decision making seems to be uncontested. On the premise that in the high-quality democracy citizens (agents) not only must have equal participation rights but \emph {must want to participate in shaping decision}, as an effective measure of democracy in a two party political system we propose the percentage of the total population that actually voted in a given elections only for two major parties. Thus, we disregard not only nonvoters but also smaller parties voters whom votes will not have a substantial impact on the election and consequently they will not be in the loop, even theoretically. To describe such a system a sociophysics model based on the $S=1$ Ising model (Blume-Capel) is proposed. The measure of democracy,  $V_D$ index, as a function of inter-party conflict is analyzed.

\end{abstract}


\maketitle

\section{Introduction}
Over half of the world's countries can be considered democracies. However, the quality of democracy in particular countries may be quite different. The question is if there is a way to distinguish the quality of democracy or in other words if there is a sensible measure of democracy. There is a number available measures - indices such as: Democracy Index, Freedom House, Polity, Democracy Barometer or Vanhanen Index which try to take into account various aspects of democracy and consequently are based on many indicators and subjective assessments. 

\emph {The democracy index} is based on 60 indicators grouped in five categories: electoral process, civil liberties, political participation and culture, and ranks countries as one of types: "full democracies", "flawed democracies", "hybrid regimes" and "authoritarian regimes" (incidentally, from a physicist point of view these categories resemble the 18th centure definitions of the Fahrenheit scale points: aestus intolerabilis (blinding heat), calor ingens (great heat), aer temperatus (moderate), aer frigidus (cold), ...) . 
\emph{Freedom House}  assesses the current state of political rights and civil liberties in each state on a scale from 1 (most free) to 7 (least free) and then classified as "free", "partly free", or "not free".
\emph{Polity's} conclusions are based on an evaluation of an election, the nature of political participation, and the extent of checks on executive authority. The Polity scale ranges from -10 to 10 from "autocracies"  (-10 to -6), through anocracies (-5 to 5) to democracies (6 to 10).
\emph{The Democracy Barometer}\cite{DB} is based on the idea that one can measure the degree of fulfillment of the nine "functions" deduced from three principles: Freedom (functions: Individual liberties, Rule of Law, Public Sphere), Control (Competition, Mutual Constraints, Governmental Capability), and Equality (Transparency, Participation, Representation). DB consists of a total of 100 indicators.
\emph{The Vanhanen Index} \cite{Tatu} is based on two clearly defined quantitative indicators corresponding to the two theoretical dimensions of democratization called: "competition" and "participation".  According to the Vanhanen idea the "degree of competition" in a given political system is indicated by the electoral success of the smaller parties, and the "degree of electoral participation" is measured by the percentage of the total population that actually voted in a given elections. These two variables are taken with the same weight to construct an index of democratization (ID) - Vanhanen Index.

None of the above mentioned indices has received common acceptance, and except for the Vanhanen Index their construction is rather complicated and linked to a certain extent with policy. So, it seems to be also helpful to analyze the problem of measure of democracy by using statistical physics models or sociophysics approach \cite{Galam, Stauffer, Fortunato}.

\section{The model}

The starting point is the premise that in the high-quality democracy citizens (agents) not only must have equal participation rights, which is obvious, but also \emph {must want to participate in shaping decision}. In this paper we confine ourselves to consider two party system i.e. political system in which the electorate votes mostly  only for two major parties. So, one or the other party can win a majority in the legislature. In consequence votes given to smaller parties  have only formal meaning without a real  impact on shaping decision. The classical example of a state with the two party system is of course the U.S. where in fact all members of the parliament belong to one of the two major parties. However, more common is the two party system where two major parties dominate elections but there are third parties which have some seats in the legislature. The examples are the United Kingdom or  Poland for eight years. 

As an effective measure of democracy in the two party political system we propose the percentage of the total population that actually voted for two major parties in a given elections. Thus we divide the whole population entitled to vote into three groups: the electorate of the first party called $L$, the electorate of the second party called $C$, and the others called $F$. The latter group form: the smaller (third parties) voters which in the main vote "against" and are fully aware that their voices will not have a major impact on the practical outcome of the election, floating voters or indifferent citizens. So, the effective democracy measure $V_D$ is given by
\begin{equation} 
V_D = \frac{N_L+N_C}{N_L+N_C+N_F}, 
\end{equation}
where $N_L, N_C, N_F$ denote numbers of the particular parties voters. 

According to the sociophysics idea social behavior can be modeled in the same way that physics models natural phenomena \cite{Galam}. The most popular and useful physics model applied to describe social behavior is undoubtedly the Ising model \cite{Galam, Stauffer, SW, SW1, Fortunato}. So, in the sociophysics language we consider a group of $n$ agents (citizens), where $n=n_L+n_C+n_F$, and  $n_L, n_C, n_F$ denote initial numbers of the "L", "C", and "F"  party voters, respectively. Each of the agent has attached an Ising variable (spin) $S_i^{\alpha}$, where $\alpha = L,C,F$ and $i=1,2,....,   n_{\alpha}$. In this case the Ising variable has three possible values, and when the agent "$i$" is a voter of "$L$" party, we take $S_i = 1$, when the agent is "$C$"-voter,  $S_i =-1$, and for "$F$"-voter, $S_i = 0$. Analogously, as in the physics case we introduce coupling between two agents from the same group $K_{\alpha}  (\alpha=L,C, F)$ which is a measure of the unity of views or satisfaction to be a member of the same group. 
In a stable situation the coupling $K_F$ is negligible because usually the members of the "F" party have no common views. To distinguish creeds of the electorates of the "$L$" and "$C$" parties we introduce an external field $H_{\beta}$ ($\beta = L, C$) coupled linearly with each agent of the "$L$" and "$C$" groups. The members of the "$F$" group are not able to distinguish between "$L$" and "$C$" party, so their "creed" has to be independent of the sign $+, -$, i.e. a "field" $D$ should be coupled to $(S_i^F)^2$. Finally, confining ourselves to the one dimensional arrangement of the particular subgroup members, one has three decoupled chains described by the Hamiltonian: 

\begin{eqnarray}
\label{1} 
\tilde H_0 &=&-\sum_{\alpha=L,C,F}K_{\alpha}\sum_{ i=1}^{n_{\alpha}} S^{\alpha}_i S^{\alpha}_{i+1}-\sum_{\beta=L,C} H_{\beta} \sum_{i=1}^{n_{\beta} }S^{\beta}_i  - D_F\sum_{i=1}^{n_F} (S_i^F)^2. 
\end{eqnarray}

Postulating a principle of maximum satisfaction \cite{Galam} one can find the equilibrium  state of the model described by the Hamiltonian (2). And if  
\begin{equation}
sgn(H_L) = - sgn(H_C),  \quad K_{\alpha}  > 0,  \quad and \quad D_F < - K_F,
\end{equation}
then, for the isolated system at the counterpart of the physical ground state (zero temperature) all agents of the "$L$" group have spin $+1$, all agents of the "$C$" group have spin $-1$, and all members of the "$F$" group have spin $0$. In this paper we consider only equilibrium properties of the system which in physics depend on temperature. In principle such a quantity does not exist in social system. However, there is the social meaning of temperature T in sociophysics as an overall approximation for all random events which influence decisions but are not included in the model \cite{Stauffer1}.  Accordingly, one can assume that social systems have their "temperature" at their steady state which validates an application of the finite temperature statistical physics methods to study social systems . Whereas, at $T=0$ all agents from $"L"$ group have spin $+1$, from $"C"$ group  $-1$, and from $"F"$ group $0$, at finite temperature only $N_{\alpha}^+$ members of $"\alpha"$ group have still spin $+1$, $N_{\alpha}^-$ spin $-1$, and $N_{\alpha}^0$ spin $0$. Consequently, at finite temperature $"\alpha"$ party has $N_{\alpha}$   voters ($\alpha = L, C, F$)
\begin{equation}
N_L =\sum_{\alpha=L,C,F} N_{\alpha}^+ ,\quad N_C =\sum_{\alpha=L,C,F} N_{\alpha}^-,\quad N_F = \sum_{\alpha=L,C,F} N_{\alpha}^0
\end{equation}
and the quantities $N_{\alpha}^{\chi}$ ($\chi = +, -, 0$) expressed by the spin averages in the following way:
\begin{equation}
N_{\alpha}^+ = \frac{1}{2}(<S_{\alpha}^2>+<S_{\alpha}>),\quad N_{\alpha}^- = \frac{1}{2}(<S_{\alpha}^2>-<S_{\alpha}>),\quad N_{\alpha}^0 = 1 - N_{\alpha}^+ - N_{\alpha}^-.
\end{equation}

As a measure of political strife between electorates of the two major parties $"L"$ and $"C"$ we introduce a coupling $Q$
\begin{equation}
 - Q \sum_i (S_i^{L} )^2(S_i^{C})^2.
\end{equation}
The choice of such a coupling prefers an exchange of voters between the $"L"$ (or $"C"$) and $"F"$ group rather than between $"L"$ and $"C"$ which is possible but less probable from the ideological point of view.

\section{The method}

The obvious way to analyze flows of the voters between the parties are computer simulations. However, in this paper we concentrate on the equilibrium properties using to study the Hamiltonian (2,6) the linear renormalization group transformation.  We start with the three decoupled chains (2), assuming that initially the number of the voters in each group $(L,C,F)$ is the same $n_{\alpha} = n$, and

\begin{eqnarray}
\label{1} 
H_0 =-\beta \tilde H_0 = \sum_{\alpha=L,C,F} H_0^{\alpha}, \qquad H_0^{\alpha}&=&k_{\alpha}\sum_{ i=1}^{n} S^{{\alpha}}_i S^{{\alpha}}_{i+1}+ h_{{\alpha}} \sum_{i=1}^{n} S^{{\alpha}}_i+d_{\alpha} \sum_{i=1}^n (S_i^{\alpha})^2 ,
\end{eqnarray}
where $k_{\alpha} = -K_{\alpha}/T,  h_{\alpha} = -H_{\alpha}/T,  d_{\alpha} = -D_{\alpha}/T$. The minimal set of the parameters to describe our model consists three intrachain couplings $k=k_L=k_C, h=h_L=-h_C$, and $d=d_F$ and yields
\begin{eqnarray}
H_0^{L}=k\sum_{ i=1}^{n} S^{{L}}_i S^{{L}}_{i+1}+ h \sum_{i=1}^{n} S^{{L}}_i, \quad
H_0^C=k\sum_{ i=1}^{n} S^{C}_i S^{C}_{i+1}-h \sum_{i=1}^{n} S^{C}_i, \quad
H_0^F=d\sum_{i=1}^n (S_i^F)^2. 
\end{eqnarray}
The renormalization group transformation for the Hamiltonian (7) is defined by
\begin{equation}
exp[H'_0(\sigma)=Tr_SP(\sigma,S)exp[H_0(S)],
\end{equation}
and the weight operator $P(\sigma,S)$ which couples the original $S$ and effective $\sigma$ spins is chosen in the linear form \cite{JS}

\begin{equation}
P(\sigma,S) = \prod_i p_i =\prod_i (1-S_{2i+1}^2-\sigma_{i+1}^2 +\frac{1}{2} S_{2i+1} \sigma_{i+1}+\frac{3}{2} S_{2i+1}^2 \sigma_{i+1}^2)
\end{equation}

For the decoupled chains the transformation (9, 10)  is a decimation transformation where in each step of the procedure every other spin is killed and the renormalized Hamiltonian can be written in the form
\begin{equation}
H'({\sigma^{\alpha}}) = \sum_{\alpha=L,C,F}\ln Tr_{S^{\alpha}} e^{H_0^{\alpha} (S^{\alpha})}
\end{equation}
Unlike the case of the two-state model ($S=1/2$ Ising model), the decimation transformation for three-state ($S=1$) model generates new interactions
\begin{equation}
j_{\alpha} (S_i^{\alpha})^2 S_{i+1}^{\alpha}, \quad q_{\alpha} (S_i^{\alpha})^2 (S_{i+1}^{\alpha})^2, \quad h_F S_i^F, \quad d_L (S_i^L)^2, \quad d_C (S_i^C)^2,
\end{equation}
and finally
\begin{eqnarray}
\ln Tr_{S^{\alpha}}  e^{H_0^{\alpha} (S^{\alpha})} & =&\ln[f_0^{\alpha} 
+ f_1^{\alpha}\sigma_i^{\alpha}
+f_2^{\alpha} \sigma_i^{\alpha} \sigma_{i+1}^{\alpha}
+ f_3^{\alpha}(\sigma_i^{\alpha})^2
+f_4^{\alpha}  (\sigma_i^{\alpha})^2 \sigma_{i+1}^{\alpha}
+f_5^{\alpha} (\sigma_i^{\alpha} \sigma_{i+1}^{\alpha})^2] \\ \nonumber
&=& z_{\alpha}+h'_{\alpha} \sigma_i^{\alpha}
+k'_{\alpha} \sigma_i^{\alpha} \sigma_{i+1}^{\alpha}
+ d'_{\alpha}(\sigma_i^{\alpha})^2
+j'_{\alpha}  (\sigma_i^{\alpha})^2 \sigma_{i+1}^{\alpha}
+q'_{\alpha} (\sigma_i^{\alpha} \sigma_{i+1}^{\alpha})^2].
\end{eqnarray}
The renormalized parameters $h'_{\alpha}, k'_{\alpha}, d'_{\alpha}, j'_{\alpha}, q'_{\alpha}$ and $z_{\alpha}$ as functions of the original interactions are presented in the Appendix A.
The constant term $z_{\alpha}$ (independent  of effective spins $\sigma$) can be used to calculate the "free energy" per site
\begin{equation}
f = \frac{1}{3}\sum_{n=1}^{\infty}\frac{z_L^{(n)}+z_C^{(n)}+z_F^{(n)}}{2^n},
\end{equation}
where $n$ numbers the RG steps, and hence the spin averages $<S^{\alpha)}>$ and $<(S^{\alpha)})^2>$.
\begin{figure}
\label{Fig_1}
 \epsfxsize=15cm \epsfbox{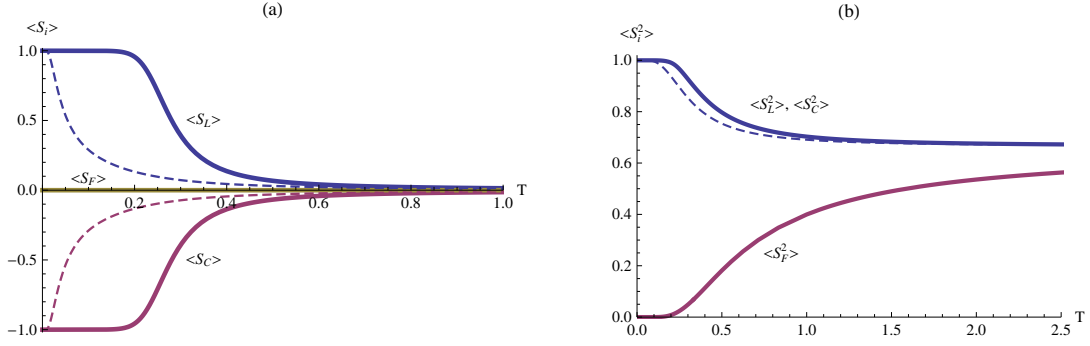}
 \caption{Temperature dependence of the magnetization  $<S_{\alpha}>$ (a) and $<S_{\alpha}^2>$ (b) for three noninteracting chains. Solid lines denote infinite chains and dashed lines three-spin chains.}
\end{figure}
In Fig.1 the temperature dependences of the spin averages found from the RG procedure for infinite chains (solid lines), and exact results for three three-site chains (dashed lines) are presented  for the model with 
\begin{equation}
 D_F = -1.1, \quad K_L = K_C = 0.5, \quad H_L = - H_C = 0.01. 
\end{equation}
As seen, the results for the infinite and three site chains converge at $T=0$ and for high temperatures.

In order to consider the interchain (intergroup) coupling (6) we apply a cluster approximation. In this approximation one considers a finite number of isolated cells (cluster) disregarding the remaining cells of the system \cite{NvL}.  Outwardly, in our case the simplest cluster possible is made of two three-site cells from $"L"$ and $"C"$ subsystems and the contribution to the renormalized energy of this cluster  is

\begin{equation}
\ln<e^{\sum_{i=1}^{3} q (S_i^L)^2 (S_i^C)^2}>_0, \qquad q = q_{LC} = -Q/T,
\end{equation}
where
\begin{equation}
<A>_0 = \frac {Tr_S A P(\sigma, S) e^H_0}{Tr_S P(\sigma, S) e^H_0}.
\end{equation}
However, as usual the RG procedure generates new couplings, whose original values are equal to zero, and one has to consider general interaction of the isolated set of the three three-site cells from $"L"$, $"C"$, and $"F"$ subsystems
\begin{equation}
\ln<e^{H_I}>
\end{equation}
with
\begin{eqnarray}
H_I &=&\sum_{\alpha \neq \beta =L,C,F}  k_{\alpha \beta} \sum_{i=1}^3 S_i^{\alpha} S_i^{\beta} + \sum_{\alpha \neq \beta =L,C,F}  q_{\alpha \beta} \sum_{i=1}^3 (S_i^{\alpha} S_i^{\beta})^2+\sum_{\alpha \neq \beta =L,C,F}  j_{\alpha \beta} \sum_{i=1}^3 (S_i^{\alpha})^2 S_i^{\beta} \\ \nonumber
&+&\sum_{\alpha \neq \beta =L,C,F}  k^{d}_{\alpha \beta} \sum_{i=1}^3 S_i^{\alpha} S_{i+1}^{\beta} + \sum_{\alpha \neq \beta =L,C,F}  q^{d}_{\alpha \beta} \sum_{i=1}^3 (S_i^{\alpha} S_{i+1}^{\beta})^2+\sum_{\alpha \neq \beta =L,C,F}  j^{d}_{\alpha \beta} \sum_{i=1}^3 (S_i^{\alpha})^2 S_{i+1}^{\beta}\\ \nonumber
&+&\sum_{\alpha \neq \beta =L,C,F}  j_{\beta \alpha } \sum_{i=1}^3 (S_i^{\beta})^2 S_i^{\alpha}+\sum_{\alpha \neq \beta =L,C,F}  j_{\beta \alpha } \sum_{i=1}^3 (S_i^{\beta})^2 S_{i+1}^{\alpha}.
\end{eqnarray}
Altogether, one has to consider, formally, 39 coupling parameters 15 single-chain (3 chains times 5 parameters) in that, in the nimimum set, three original $k, h, d$ (Eq.8) and 24 interchain couplings (Eq.19) in that one original $q$ (Eq.16). However, it is quite easy to find analytical forms of the renormalized couplings and perform the RG iterations.
\section{The $V_D$ index}

As mentioned in the Introduction the random events but also an information noise can play in social systems a similar role as temperature in physical systems. So, one can find the temperature dependences of the number of particular party voters and $V_D$. We start with the noninteracting subsystems models: (i) defined by the parameters (15) and to check the possible role of the $K_F$ coupling (ii) additionally $K_F=1$. 
Because we assume that numbers of each group voters are equal to each others $n_L=n_C=n_F$, hence initially (at the ground state)  the numbers of each group $"L"$, $"C"$ and $"F"$ voters are the same $N_L=N_C=N_F=1$ per agent.
\begin{figure}
\label{Fig_2}
 \epsfxsize=15cm \epsfbox{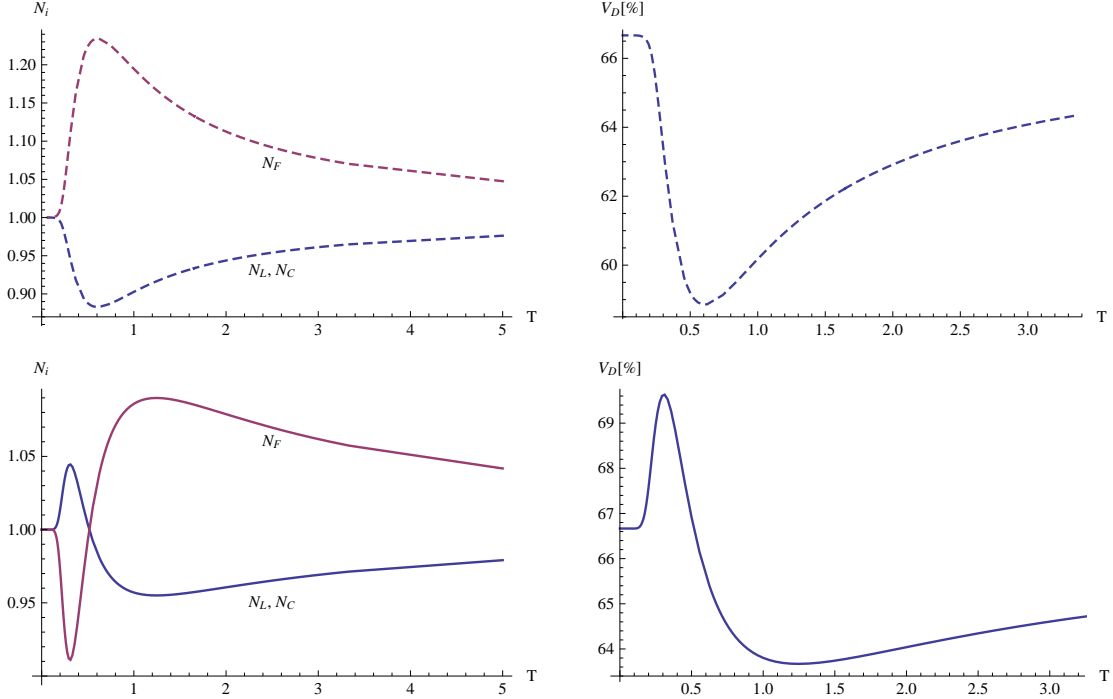}
 \caption{ The temperature dependences of the voter numbers $N_i$ per agent and index $V_D$ (in percent) for $ K_C=K_L=0.5, K_F=0$ (dashed lines) and $K_C=K_L=0.5, K_F=1,$ (solid lines).}
\end{figure}

The results are presented in Fig.2. As seen for model (i) the voter numbers start with $N_i=1$ and then with the temperature increasing, $N_L$ and $N_C$ decrease, reach a minimum and then increase to $1$ at $T ->\infty$. Similarly, the $V_D$-index starts with $\frac{2}{3}$ at $T=0$ and passing the  minimum reaches the same value $\frac{2}{3}$ at $T ->\infty$. For the model (ii) both $N_L (N_C)$ and $V_D$ first increase, reach the maxima, pass trough a minima, and then increase again to $1$ and $\frac{2}{3}$ at $T ->\infty$, respectively.

We are now in a position to evaluate the dependence of the particular parties voter numbers on the intergroup coupling $Q$ (6, 16). Let us first consider the finite system of three chains of three agents in each of them with the coupling parameters as in (15) and $Q>0$ at $T=0$ (ground state), $T=0.05$, and $T=0.2$. As is seen from the left plots of Fig.3 at the ground state the initial agent arrangement $N_L=N_C=N_F=1$ is conserved until $Q=\frac{1}{3}$, then there is a jump of $N_L=N_C$ to $\frac{1}{2}$, and $N_F$ to $2$. The jump is gradually smeared by rising temperature (middle and right plots of Fig.3). For a nonsymmetric case $K_L \neq K_C$ (Fig.4), the ground state configuration is essentially different and if, for example $K_L > K_C$ then $N_L$ does not depend on $Q$ whereas $N_C$ drops to $0$ and $N_F$ jumps to $2$ at $Q=\frac{1}{3}$. At higher temperature the behavior of $N_i$ in the nonsymmetric case is similar to that of symmetric one.  In the bottom plots of Figs.3 and 4 the Q-dependence of the $V_D$ index is shown. For low temperature $V_D$ drops sharply at $Q=\frac{1}{3}$, and for higher temperature it decreases rapidly from $\frac{2}{3}$ to $\frac{1}{3}$.
 
Now we proceed to the RG analysis of the infinite chains. To calculate the average (18) we use the identity
\begin{eqnarray}
\exp [k_{\alpha \beta} S^{\alpha} S^{\beta}+q_{\alpha \beta} (S^{\alpha} S^{\beta})^2+j_{\alpha \beta} (S^{\alpha})^2 S^{\beta}+j_{\beta \alpha} S^{\alpha} (S^{\beta})^2] = \\ \nonumber
1 + K_{\alpha \beta} S^{\alpha} S^{\beta}+Q_{\alpha \beta} (S^{\alpha} S^{\beta})^2+J_{\alpha \beta} (S^{\alpha})^2 S^{\beta}+J_{\beta \alpha} S^{\alpha} (S^{\beta})^2,
\end{eqnarray}
where
\begin{eqnarray}
K_{\alpha \beta} &=& \frac{1}{4}e^{q_{\alpha \beta}-k_{\alpha \beta}-j_{\alpha \beta}-j_{\beta \alpha}}(e^{2k_{\alpha \beta}}-e^{2j_{\alpha \beta}}-e^{2k_{\beta \alpha}+2j_{\alpha \beta}+2j_{\beta \alpha}}), \\ \nonumber
Q_{\alpha \beta} &=& -1+\frac{1}{4}e^{q_{\alpha \beta}+k_{\alpha \beta}-j_{\alpha \beta}-j_{\beta \alpha}}+\frac{1}{4}e^{q_{\alpha \beta}-k_{\alpha \beta}+j_{\alpha \beta}-j_{\beta \alpha}}+\frac{1}{4}e^{q_{\alpha \beta}-k_{\alpha \beta}-j_{\alpha \beta}+j_{\beta \alpha}}+\frac{1}{4}e^{q_{\alpha \beta}+k_{\alpha \beta}+j_{\alpha \beta}+j_{\beta \alpha}}, \\ \nonumber
J_{\alpha \beta} &=& -\frac{1}{4}e^{q_{\alpha \beta}+k_{\alpha \beta}-j_{\alpha \beta}-j_{\beta \alpha}}+\frac{1}{4}e^{q_{\alpha \beta}-k_{\alpha \beta}+j_{\alpha \beta}-j_{\beta \alpha}}+\frac{1}{4}e^{q_{\alpha \beta}-k_{\alpha \beta}-j_{\alpha \beta}+j_{\beta \alpha}}+\frac{1}{4}e^{q_{\alpha \beta}+k_{\alpha \beta}+j_{\alpha \beta}+j_{\beta \alpha}}, \\ \nonumber
J_{\beta \alpha} &=& -\frac{1}{4}e^{q_{\alpha \beta}+k_{\alpha \beta}-j_{\alpha \beta}-j_{\beta \alpha}}-\frac{1}{4}e^{q_{\alpha \beta}-k_{\alpha \beta}+j_{\alpha \beta}-j_{\beta \alpha}}+\frac{1}{4}e^{q_{\alpha \beta}-k_{\alpha \beta}-j_{\alpha \beta}+j_{\beta \alpha}}+\frac{1}{4}e^{q_{\alpha \beta}+k_{\alpha \beta}+j_{\alpha \beta}+j_{\beta \alpha}}.
\end{eqnarray}
To evaluate the RG transformation one has to know the chain averages $<S_i^{\alpha}>,  <(S_i^{\alpha})^2>$, $<S_i^{\alpha}S_i^{\beta}>, <(S_i^{\alpha}S_i^{\beta})^2>$, and $<(S_i^{\alpha})^2S_i^{\beta}>$.  It is quite easy to find their closed expressions and for example:
\begin{eqnarray}
<S_1^{\alpha}> &=& \sigma_1^{\alpha}, \quad <S_3^{\alpha}> = \sigma_2^{\alpha}, \quad \alpha = L,C,F \\ \nonumber
<S_2^{\alpha}> & =& G_0^{\alpha} + G_1^{\alpha} (\sigma_1^{\alpha}+\sigma_2^{\alpha}) + g_g^{\alpha} \sigma_1^{\alpha} \sigma_2^{\alpha} + G_2^{\alpha} [(\sigma_1^{\alpha})^2+(\sigma_2^{\alpha})^2] \\ \nonumber
&+& G_3^{\alpha} [(\sigma_1^{\alpha})^2 \sigma_2^{\alpha} +(\sigma_2^{\alpha})^2 \sigma_1^{\alpha}]  + G_4^{\alpha} (\sigma_1^{\alpha})^2 (\sigma_2^{\alpha})^2.
\end{eqnarray}
The coefficients $G_i^{\alpha}$ are presented in Appendix B.

\begin{figure}
\label{Fig_3}
 \epsfxsize=17cm \epsfbox{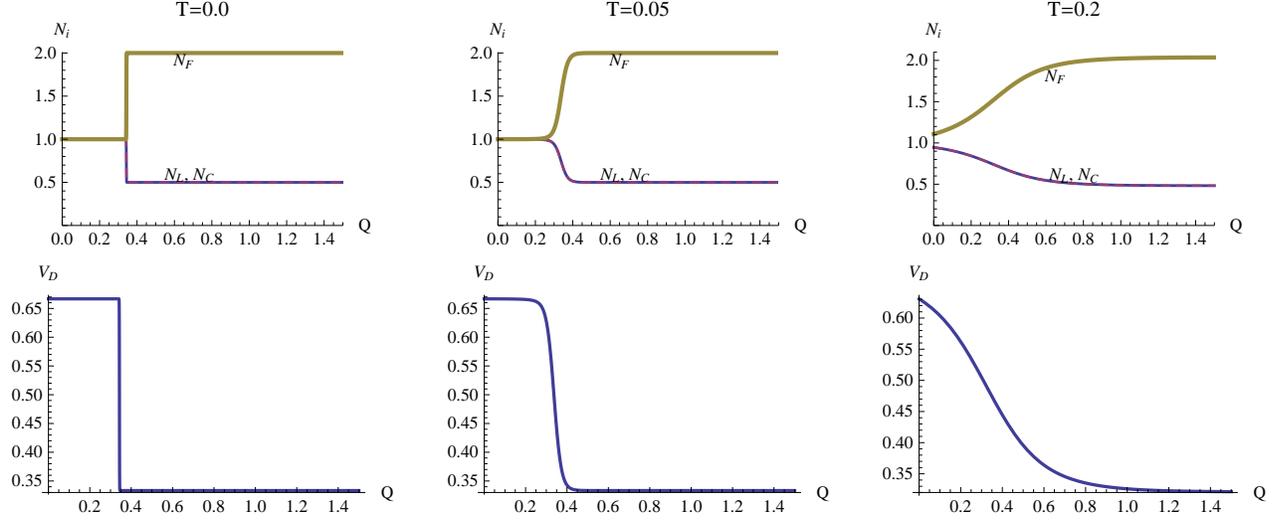}
 \caption{ Finite system:  Q-dependence of $N_i$ for $K_L=K_C=0.5$ and several temperatures.}
\end{figure}
In the cluster approximation with three three-site (-agent) blocks $"L", "C", "F"$ taking into account only two-site coupling, the RG transformation has the form of 39 recursion relations. Iterating these relations and collecting the constant terms generated in each step of the iteration process one can calculate numerically the "free energy" and then the averages $<S_i^{\alpha}>$ and  $<(S_i^{\alpha})^2>$. In Fig.5 these averages are presented as functions of interblock coupling $Q$ for the model with $D_F=-1.1, H_L=-H_C=0.01$ in two cases: (i) symmetric $K_L=K_C=0.5$ and (ii) nonsymmetric $K_L=0.5$ and $K_C=0.48$ at $T=0.25$. Knowing the averages $<S_i^{\alpha}>$ and  $<(S_i^{\alpha})^2>$ one can find the number of particular parties voters (5) and $V_D$ index (1). The results for symmetric and nonsymmetric cases are  presented in Figs. 6 and 7, respectively. As seen the dependences of the voter numbers on $Q$ for infinite system differ significantly from those for three-site blocks. However, in both cases symmetric and nonsymmetric as for the finite system at low temperature the $V_D$ index changes slowly for sufficiently small $Q$ and then drop sharply to a constant value.
    
In physical systems the coupling parameters $K_{\alpha}, D_{\alpha}, H_{\alpha}$ or $Q_{\alpha}$ have plausible interpretation even if they have an effective character. Such an interpretation is not of course so obvious for social systems.  However, one can assume that there is a positive coupling between the members of the same political environment which measure is the parameter  $K_{\alpha}$  and some parameter which separates the creeds of the particular party voters $H_{\alpha}$ . Analogously, a negative $D_{F}$ can be considered as a measure of discouragement to take part in public life and on the other hand, a positive $D_{L(C)}$  is a measure of citizen participation. In Fig.8 the $Q$-dependences of index $V_D$  of the symmetric ($K_L=K_C=0.5, D_F=-1.1, D_{L(C)}=0$) and nonsymmetric ($K_L=0.5, K_C=0.4, D_F=-1.1, D_{L(C)}=0$) models considered above are compared with the results for the model with $K_F=1$ and positive $D_{L(C)}=0.5$. As one would expect in the latter case the range of $Q$ in which $V_D$ changes ever so slightly is much broader. 

\begin{figure}
\label{Fig_4}
 \epsfxsize=17cm \epsfbox{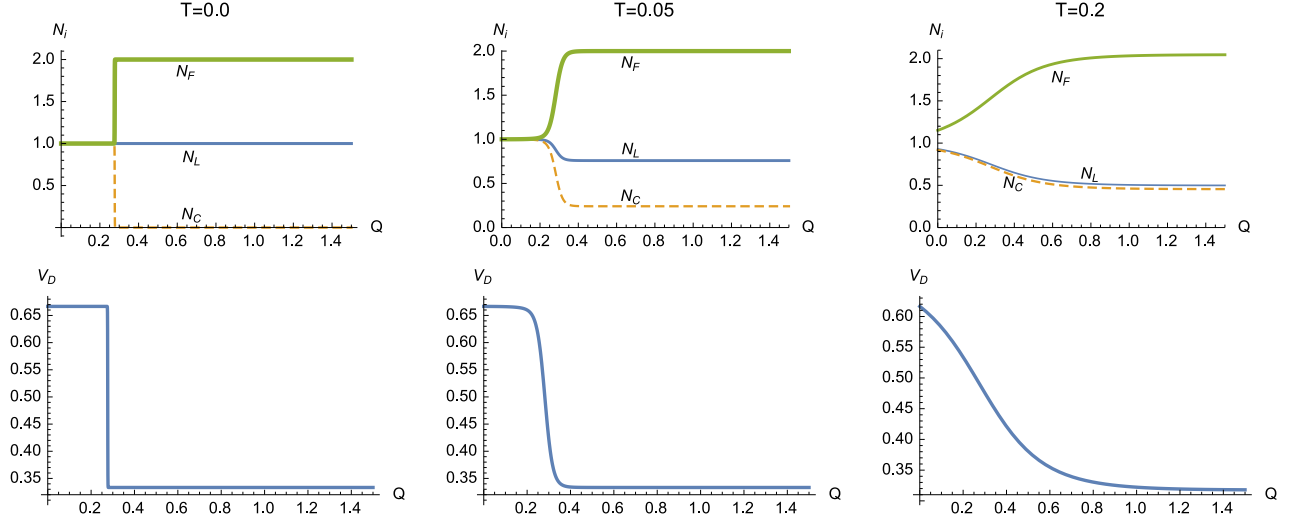}
 \caption{ Finite system:  Q-dependence of $N_i$ for $K_L=0.5,K_C=0.48$ and several temperatures.}
\end{figure}

\begin{figure}
\label{Fig_5}
 \epsfxsize=15cm \epsfbox{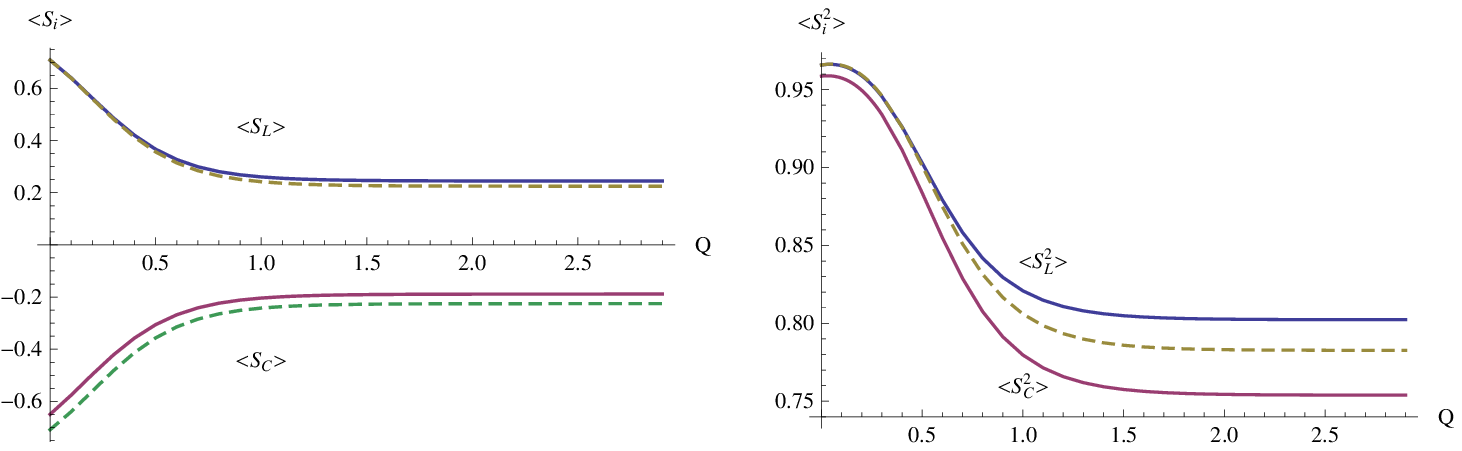}
 \caption{Infiinite chains: Magnetizations $<S_{\alpha}>$ and $<S_{\alpha}^2>$  for $K_L=0.5,K_C=0.48$ (solid lines) and  for $K_L = K_C=0.5$ (dashed lines) at $T=0.25$}
\end{figure}

\begin{figure}
\label{Fig_6}
 \epsfxsize=15cm \epsfbox{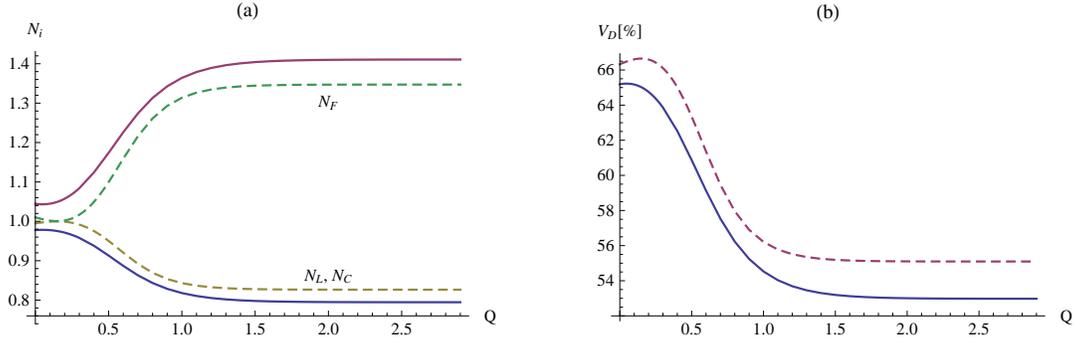}
 \caption{ Voter numbers $N_{\alpha}$ (a)  and $V_D$-index (b) as functions of $Q$ for $K_L=K_C=0.5$ at $T=0.25$ (solid lines) and  $T=0.2$ (dashed lines)}
\end{figure}

\begin{figure}
\label{Fig_7}
 \epsfxsize=15cm \epsfbox{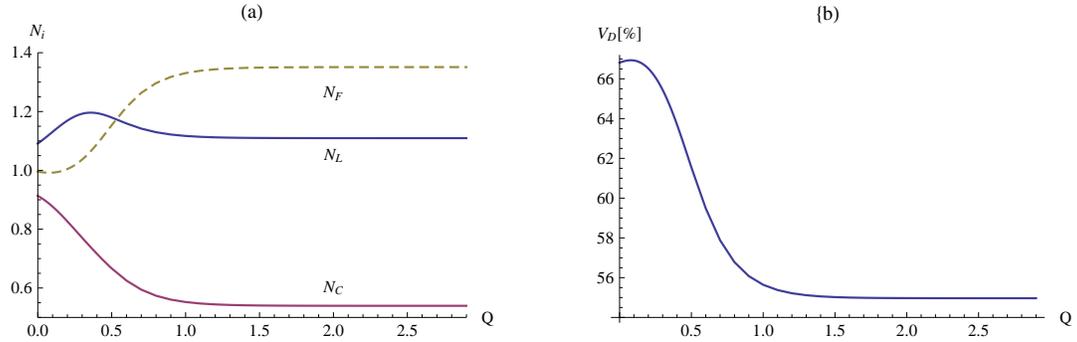}
 \caption{Voter numbers $N_{\alpha}$ (a) and $V_D$ -index (b) as functions of $Q$ for nonsymmetric model with $K_L= 0.5, K_C = 0.4, K_F=1$ at $T=0.2$.}
\end{figure}

\begin{figure}
\label{Fig_8}
 \epsfxsize=10cm \epsfbox{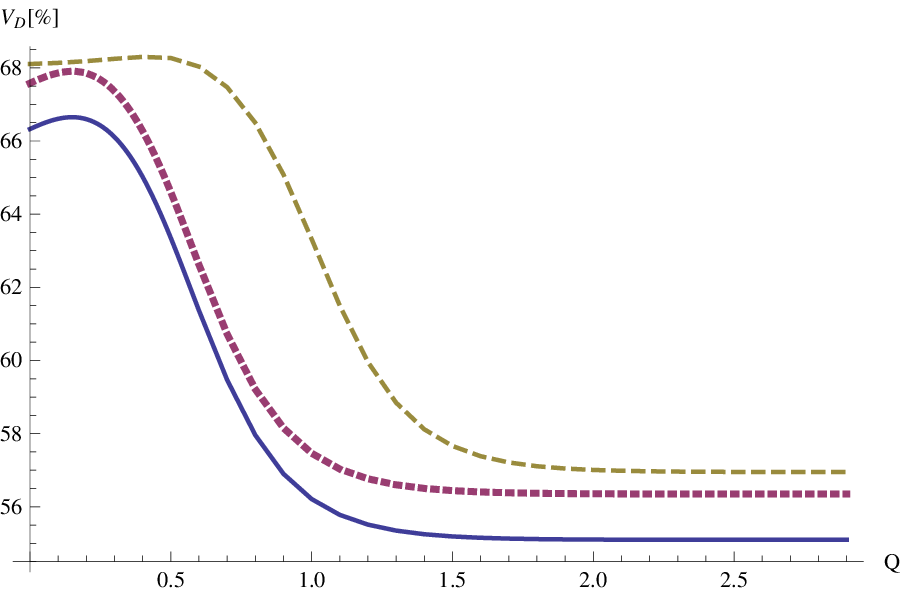}
 \caption{ $V_D$ index as functions of $Q$ for three models: symmetric $K_L=K_C= 0.5, K_F=0$ (solid line), nonsymmetric with $K_L= 0.5, K_C = 0.4, K_F=1$ (dotted), and nonsymmetric with $K_L= 0.5, K_C = 0.4, K_F=1, D_L=D_C=0.5$ (dashed) at $T=0.2$.}
\end{figure}

\section{Summary}

It is unlikely that a simple statistical physics model could be used to predict a social event, although certain sociophysicists believe that it is possible in some cases and  for example Serge Galam \cite{Galam} claims "I do not think history could be predicted even in principle, given our current tools of research and perception of the world", however, at the same time he expresses a hope that "sociophysics in the future may yield real predictive tools". Anyway, it seems that sociophysics models can be successfully used to describe, explain and point out general features of social behavior. 

In this paper to describe an influence of the social interplay between electorates of the two major parties, embodies by the coupling $Q$, on the quality of democracy,  we propose the three-state Ising-like statistical physics model. 
The minimal number of parameters which define the model is three: the measure of the unity views of the two major parties ($L, C$) voters - $k=-K_L/T=-K_C/T$,  the field which differentiates the creeds of the particular party voters - $h=-H_L/T=-H_C/T$, and the measure of a discouragement to take part in the public life of the $F$ group citizens - $d=-D_F/T$. 
The measure of the democracy quality is $V_D$, index defined as a percentage of the total population that actually voted for two major parties in a given election. This index reflects not only rights but also the inclination of the citizens to participate in decision making, even if theoretically, which can be treated as an essence of democracy. 
To check a universality of the results we have applied three sets of the original parameters: (i) symmetric model ($K_L=K_C=0.5, K_F=0, D_F=-1.1, D_{L(C)}=0$), (ii) nonsymmetric model ($K_L=0.5, K_C=0.4$, other as above), and (iii)  ($K_L=0.5, K_C=0.4, K_F=1, D_{L(C)}=0.5$, other as above). In all cases, there is a range of $Q$ in which the index $V_D$ changes slightly, first increasing with $Q$, passes a maximum then at some characteristic point $Q_f$ starts to fall rapidly, and at $Q_c$ reaches a constant value (Fig.8). 
At the same time for the symmetric model (i) the numbers of both major parties voters firstly slightly increase with increasing $Q$ and then sharply decrease (Fig.6). For the nonsymmetric case ($K_L>K_C$), only the number of $L$ party voters increases, reaches a maximum and then drops to some constant value whereas the number of $C$ party voters decreases immediately with increasing of $Q$ (Fig.7). When the value of both $Q_f$, $Q_c$, and the location of the $V_D$ maximum depend on the model parameters, a  collapse of $V_D$ seems to be a general feature of the present model. 

We conclude from the model that in the two party political system a reasonable conflict level between the electorates of the two major parties can be mutually beneficial for both parties and what is more for the quality of democracy measured by the index $V_D$. However, for the higher conflict level (higher degree of polarization), citizen participation decreases rapidly. For $Q > Q_c$ only so called hard or fixed electorates of the major parties want in public life. High percentage, and in the extreme case most of the society, decline voting for a party which can win a majority in the legislature ergo decline participation in a real decision making, which in fact means the collapse of the high quality democracy.
\section{Appendix}
\subsection{Decimation transformation parameters}
The renormalized parameters (13) as functions of the original interactions $k_{\alpha},  h_{\alpha}, d_{\alpha} $(RG recursion relations).
\begin{eqnarray}
z_{\alpha} &=& \lambda_0^{\alpha}, \quad  h'_{\alpha} = \lambda_1^{\alpha} - \lambda_2^{\alpha},\quad  
k'_{\alpha} = \frac{1}{4}(-2 \lambda_3^{\alpha} + \lambda_4^{\alpha} + \lambda_5^{\alpha}) , \quad
d'_{\alpha} = -2 \lambda_0^{\alpha} + \lambda_1^{\alpha} + \lambda_2^{\alpha}) ,\\ \nonumber
j'_{\alpha} &=& \frac{1}{4}(-2 \lambda_1^{\alpha} + 2\lambda_2^{\alpha} + \lambda_4^{\alpha} +\lambda_5^{\alpha}) , \quad 
q'_{\alpha} = \frac{1}{4}(4 \lambda_0^{\alpha} -4 \lambda_1^{\alpha} - 4 \lambda_2^{\alpha} +2 \lambda_3^{\alpha}+ \lambda_4^{\alpha} +\lambda_5^{\alpha}). 
\end{eqnarray}

\begin{equation}
\lambda_i^{\alpha} = \ln f_i^{\alpha}, \quad \omega_i^{\alpha} = \frac{1}{f_i^{\alpha}}, \quad  i = 0,1,..,5, \quad \alpha = L, C, F.
\end{equation}

\begin{eqnarray}
f_0^{\alpha} &=& 1 + e^{d_{\alpha} - h_{\alpha}} + e^{d_{\alpha} + h_{\alpha}}, \\ \nonumber
f_1^{\alpha} &=&  e^{\frac{1}{2}(d_{\alpha} - h_{\alpha} - 2k_{\alpha})} ( e^{d_{\alpha} + q_{\alpha}} +e^{h_{\alpha} + k_{\alpha}} +e^{d_{\alpha} +2 h_{\alpha}+q_{\alpha}+2 k_{\alpha} +2 j_{\alpha}}) ,  \\ \nonumber
f_2^{\alpha} &=& e^{\frac{1}{2}(d_{\alpha} - 3h_{\alpha} + 2k_{\alpha}+4 j_{\alpha})} 
( e^{d_{\alpha} + q_{\alpha} + 2 k_{\alpha}} + e^{d_{\alpha} + 2 h_{\alpha}+q_{\alpha} + 2 j_{\alpha}} 
+e^{h_{\alpha} +k_{\alpha}+2 j_{\alpha}}) ,  \\ \nonumber
f_3^{\alpha} &=& e^{d_{\alpha}} 
(1 + e^{d_{\alpha} - h_{\alpha} +2 q_{\alpha} -2 j_{\alpha}} +e^{d_{\alpha} + h_{\alpha}+2 (q_{\alpha}+j_{\alpha}}) , \\ \nonumber
f_4^{\alpha} &=& e^{d_{\alpha}} 
( e^{h_{\alpha}} + e^{d_{\alpha} + 2 q_{\alpha}-2 k_{\alpha}} +e^{d_{\alpha} + 2( h_{\alpha} + q_{\alpha} + k_{\alpha}+2 j_{\alpha}}  ) , \\ \nonumber
f_5^{\alpha} &=& e^{d_{\alpha}} 
( e^{-h_{\alpha}} + e^{d_{\alpha} + 2 q_{\alpha}-2 k_{\alpha}} +e^{d_{\alpha} + 2(-h_{\alpha} + q_{\alpha} + k_{\alpha} - 2 j_{\alpha}}  ) .
\end{eqnarray}

\subsection{Single chain averages}
\begin{eqnarray}
G_0^{\alpha} &=& c_p^{\alpha} g_0^{\alpha}, \quad G_1^{\alpha} = (c_p^{\alpha}+c_d^{\alpha}) g_1^{\alpha} + c_h^{\alpha} (g_0^{\alpha} + g_2^{\alpha}), \quad
G_2^{\alpha} = c_h^{\alpha} g_1^{\alpha}+c_p^{\alpha} g_2^{\alpha}+c_d^{\alpha} (g_0^{\alpha}+g_2^{\alpha}),
\\ \nonumber
G_g^{\alpha} &=& 2(c_q^{\alpha}+c_h^{\alpha})(g_1^{\alpha}+g_3^{\alpha})+c_k^{\alpha}(g_0^{\alpha}+2 g_2^{\alpha}+g_4^{\alpha})+(2c_d^{\alpha}+c_j^{\alpha}+c_p^{\alpha})g_g^{\alpha}, \\ \nonumber
G_3^{\alpha} &=& (c_j^{\alpha}+c_k^{\alpha}) g_1^{\alpha} + c_h^{\alpha} (g_2^{\alpha}+g_4^{\alpha}+g_g^{\alpha})+(c_j^{\alpha}+cK-^{\alpha}+c_p^{\alpha})  g_3^{\alpha}+c_d^{\alpha}(g_1^{\alpha}+2 g_3^{\alpha}) \\ \nonumber
&+&c_q^{\alpha}(g_0^{\alpha}+2 g_2^{\alpha}+g_4^{\alpha}+g_g^{\alpha}), \\ \nonumber
G_4^{\alpha} &=& 2c_d^{\alpha}(g_2^{\alpha}+g_4^{\alpha})+2 c_h^{\alpha} g_3^{\alpha}+2c_q^{\alpha}(g_1^{\alpha}+g_3^{\alpha})+c_p^{\alpha} g_4^{\alpha}+c_j^{\alpha}(g_0^{\alpha}+2 g_2^{\alpha}+g_4^{\alpha})+c_k^{\alpha} g_g^{\alpha}.
\end{eqnarray}
where
\begin{eqnarray}
c_p^{\alpha}&=&\omega_0^{\alpha}, \quad c_h^{\alpha}=\frac{1}{2}(\omega_1^{\alpha}-\omega_2^{\alpha}), \quad c_k^{\alpha} = \frac{1}{4}(\omega_5^{\alpha}+\omega_4^{\alpha}-2 \omega_3^{\alpha}), \quad c_d^{\alpha}=\frac{1}{2}(\omega_1^{\alpha}+\omega_2^{\alpha})-\omega_0^{\alpha}, \\ \nonumber
c_q^{\alpha}&=&\frac{1}{2}(\omega_2^{\alpha}-\omega_1^{\alpha}+\omega_4^{\alpha}-\omega_5^{\alpha}), \quad
c_j^{\alpha}=\omega_0^{\alpha}-\omega_1^{\alpha}-\omega_2^{\alpha}+\frac{1}{2}(\omega_3^{\alpha}+\omega_4^{\alpha}+\omega_5^{\alpha}).
\end{eqnarray}
and
\begin{eqnarray}
g_0^{\alpha} &=& Tr_S S_2^{\alpha}[1 - (S_1^{\alpha})^2] [1 - (S_3^{\alpha})^2]e^{H_0^{\alpha}}, \quad 
g_1^{\alpha} = \frac{1}{2} Tr_S S_2^{\alpha} S_1^{\alpha}) [1 - (S_3^{\alpha})^2]e^{H_0^{\alpha}}, \\ \nonumber
g_g^{\alpha} &=& \frac{1}{4}Tr_S S_2^{\alpha}S_1^{\alpha}S_2^{\alpha}e^{H_0^{\alpha}}, \quad 
g_2^{\alpha} = Tr_S S_2^{\alpha} [-1+\frac{3}{2} (S_1^{\alpha})^2][1-S_3^{\alpha})^2) e^{H_0^{\alpha}},\\ \nonumber
g_3^{\alpha} & =& \frac{1}{2}Tr_S S_2^{\alpha} S_3^{\alpha}[-1+\frac{3}{2} (S_1^{\alpha})^2] e^{H_0^{\alpha}}, \quad
g_4^{\alpha} = Tr_S S_2^{\alpha}[-1+\frac{3}{2} (S_1^{\alpha})^2][-1+\frac{3}{2} (S_3^{\alpha})^2] e^{H_0^{\alpha}}.
\end{eqnarray}


\end{document}